\documentclass[preprint,proceedings]{rmaa}
\usepackage{psfrag,color}
\usepackage{epsf}

\newcommand{\ha}{{\rm H$\alpha$ }}

\newcommand{\hans}{{\rm H$\alpha$}}
\newcommand{\hii}{{\rm H\,{\sc ii} }}

\newcommand{\fuv}{{\rm FUV }}
\newcommand{\fuvns}{{\rm FUV}}

\newcommand{\pegase}{{\sc P\'egase }}

\newcommand{\afuv}{{$A_{\rm FUV}$ }}
\newcommand{\afuvns}{{$A_{\rm FUV}$}}

\SetYear{2002}
\SetConfTitle{Galaxy Evolution: Theory and Observations}

\title{Dust-induced Systematic Errors in Ultraviolet-Derived Star
Formation Rates} 

\author{
  Eric F. Bell\affil{Steward Observatory, Tucson AZ, USA.}
}

\shortauthor{Bell}
\shorttitle{Systematic Errors in UV-derived SFRs}

\fulladdresses{
\item Eric F. Bell: Steward Observatory, University of Arizona,
	933 North Cherry Avenue, Tucson AZ, 85716, USA (ebell@as.arizona.edu).
}

\listofauthors{Eric F. Bell}
\indexauthor{Bell, E. F.}

\abstract{Rest-frame far-ultraviolet (FUV) luminosities form the 
`backbone' of our understanding of star formation at all cosmic epochs.
FUV luminosities are typically corrected for dust by assuming
that extinction indicators which have been calibrated for local
starbursting galaxies apply to all star-forming galaxies.  
I present evidence that `normal' star-forming galaxies
have systematically redder UV/optical colors than starbursting
galaxies at a given FUV extinction.  This is attributed
to differences in star/dust geometry, coupled with a small
contribution from older stellar populations.  Folding in data for 
starbursts and ultra-luminous infrared galaxies, I conclude
that SF rates from rest-frame UV and optical data alone are 
subject to large (factors of at least a few) systematic uncertainties
because of dust, which cannot be reliably corrected for
using only UV/optical diagnostics.
}

\addkeyword{Dust, Extinction}
\addkeyword{Galaxies: General}
\addkeyword{Galaxies: Stellar Content}
\addkeyword{Ultraviolet: Galaxies}

\begin{document}
% Typeset article header
\maketitle

\section{Introduction} \label{sec:intro}

Understanding the star formation (SF) rates
of galaxies, at a variety of cosmic
epochs, is a topic of intense current interest
(e.g., Yan et~al.\@ 1999; Blain et~al.\@ 1999; Haarsma et~al\@ 2000).
Many SF rates are derived from highly dust-sensitive rest frame 
far-ultraviolet (FUV) luminosities (e.g., Madau et~al.\@ 1996; Steidel
et~al\@ 1999).
In the local Universe, Calzetti et~al.\@ (1994,1995) and 
Meurer et~al.\@ (1999)
found a tight correlation between ultraviolet (UV)
spectral slope $\beta$\footnote{Defined by $F_{\lambda} 
\propto \lambda^{\beta}$, where $F_{\lambda}$ is the flux per 
unit wavelength $\lambda$.} and the
attenuation\footnote{Attenuation differs from extinction in 
that attenuation describes the amount of light lost because of dust at a given
wavelength in systems with complex star/dust geometries
where many classic methods for determining extinction, such as
color excesses, may not apply.} in the FUV (\afuvns), for a sample of  
in\-hom\-ogen\-eous\-ly-selected starburst galaxies.  
This correlation's low scatter 
requires a constant intrinsic value of $\beta \sim -2.5$
for young stellar populations (e.g., Leitherer et~al.\@ 1999), coupled
with some regularities in the distribution and extinction properties
of dust (e.g., Gordon et~al.\@ 1997).
Assuming that this
correlation holds for all galaxies at high redshift, this was used to 
correct the FUV flux for extinction in a statistical
sense (see, e.g., Adelberger \& Steidel 2000 and references therein).

However, recent work has called the universality of 
the $\beta$--\afuv correlation into question.  
Radiative transfer models predict a large scatter between
$\beta$ and \afuv (Witt \& Gordon 2000).
Furthermore, both Large Magellanic Cloud (LMC)
\hii regions (Bell et~al.\@ 2002) and ultra-luminous infrared galaxies
(ULIRGs; Goldader et~al.\@ 2002) do not obey the starburst
correlation.  Tantalizingly, there are indications that
`normal', quiescent star-forming galaxies have less 
UV extinction than predicted by the Calzetti et al.\
relation (Buat et~al.\@ 2002).  Taken together, these issues
raise serious questions about the applicability of rest-frame 
UV-derived SF rates for non-starbursting 
galaxies.

Here, I investigate the relationship between
$\beta$ and \afuv for quiescent, `normal' star-forming galaxies
for the first time (to date, this correlation has been examined
directly for starbursts, ULIRGs and \hii regions only).  
For more details, see Bell (2002).

\section{The $\beta$--\afuv correlation for normal galaxies} \label{sec:beta}

\begin{figure}[tbh]
\vspace{-0.5cm}
\hspace{-0.7cm}
\epsfxsize=9.0cm
\epsfbox{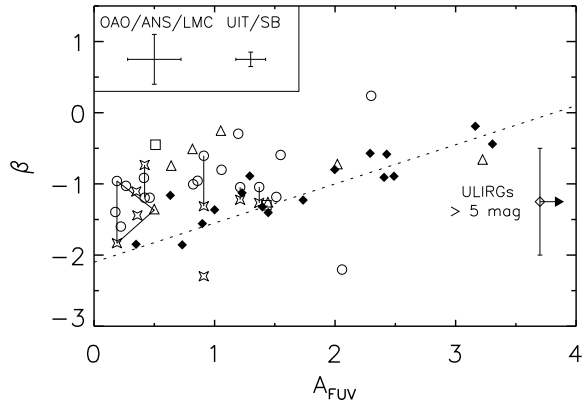}
\vspace{-0.7cm}
\caption{\label{fig:beta} 
UV spectral slope $\beta$ against attenuation at
$\sim 1500${\AA}, \afuvns.  {\it UIT} galaxies are plotted
as triangles, {\it OAO} galaxies as stars, {\it ANS} 
galaxies as circles, the LMC and 30 Dor are plotted
as squares, and starbursts are plotted as filled
diamonds.  ULIRGs (open diamond) typically have $A_{\rm FUV} \ga 5$ mag,
and `blue' $\beta$ values (Goldader et~al.\@ 2002).  
The dotted line shows a rough `by-eye' fit
to the starbursts.  
Different measurements for the same galaxies are 
connected by solid lines.
}
\end{figure}

In Fig.\ \ref{fig:beta}, I show the UV spectral slope $\beta$ 
between 1500{\AA} and 2500{\AA} for normal galaxies (calculated
using integrated and/or large-aperture photometry) against
the FUV attenuation as open symbols.  I calculate the FUV attenuation
by balancing the FUV and FIR luminosity of my sample galaxies.
Gordon et~al.\@ (2000) show that this is a robust indicator of FUV 
attenuation from a modelling perspective. Furthermore,
in Bell (2002, in preparation)
I argue, by comparing FIR$+$FUV vs.\ extinction-corrected \ha
SF rates, that FUV/FIR reflects the real FUV attenuation with much better than 
a factor of two systematic and random error.  I show $\beta$ against
\afuv for starbursting galaxies 
(solid diamonds; Calzetti et~al.\@ 1994,1995) and ULIRGs
(Goldader et~al.\@ 2002) for comparison.   Clearly, normal galaxies
have substantially redder UV spectral slopes, by $\Delta \beta \sim 1$,
than their starburst counterparts
at a given \afuv (derived using FUV/FIR energy balance).
This offset between starbursting
and normal galaxies is seen by 7 different experiments
({\it UIT}, {\it OAO}, {\it ANS}, and {\it IUE} for normal galaxies; 
a sounding rocket, {\it D2B-Aura} and {\it TD1} for the LMC).
Furthermore, normal galaxies exhibit substantially
larger scatter than the starbursts.
Interestingly, this large scatter is largely intrinsic. The galaxies
which are closer to the starburst galaxy relationship tend
to be the most vigorously star-forming members of the 
`normal' galaxy sample, and the redder ones are 
more quiescient (but still star-forming).

\section{Exploring the origins of the $\beta$--\afuv correlation} 
\label{sec:origin}

\begin{figure}[tbh]
\vspace{-0.5cm}
\hspace{-0.7cm}
\epsfxsize=9cm
\epsfbox{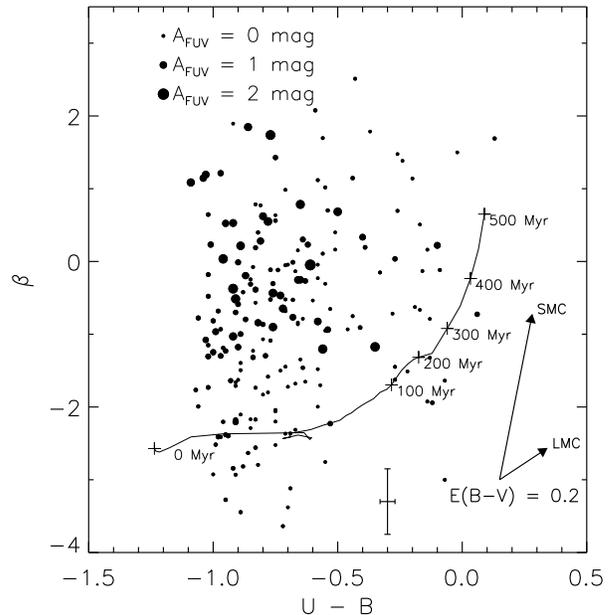}
\vspace{-0.7cm}
\caption{\label{fig:bica} 
UV spectral slope $\beta$ against $U - B$ color (a reasonable
age indicator) for a sample of 198 UV-bright stellar clusters and
associations in the LMC.  Symbols are coded by FUV attenuation:
larger symbols depict more highly attenuated clusters.  
Overplotted are the \pegase stellar population model colors
for a single burst with ages $<500$ Myr (Fioc \& Rocca-Volmerange, 
in preparation).  The effect of 
dust reddening, assuming a SMC bar-type or LMC-type dust screen,
is shown.  The dust vectors are shown simply to give  
some intuition about the effects of dust: radiative
transfer effects and/or extinction curve variations in the UV
make the prediction of the detailed effects of dust on a plot of this
type challenging.
}
\end{figure}

It is reasonable to hypothesize that stellar population
and/or dust effects contribute to the
difference in behavior between normal and starburst galaxies
on the $\beta$--\afuv plane.
Stellar population models show that older {\it star-forming}
stellar populations are somewhat redder ($\Delta \beta \la 0.5$)
than younger star-forming populations
(e.g., Leitherer et~al.\@ 1999).
Alternatively, radiative transfer models 
can easily generate relatively large changes in 
$\beta$ for only modest \afuv by appealing to different
star/dust geometries and/or extinction curves 
(e.g., Gordon et~al.\@ 2000; Bell et~al.\@ 2002).

In order to test why normal galaxies have redder $\beta$ values
than starbursts, independent constraints on a galaxy's SF
history are required (to allow splitting of age and dust
effects).  Independent age constraints are available
for stellar clusters and associations in the LMC (in the form of
$U - B$ optical color).  I construct 
matched aperture values of $\beta$ and \afuv for 
198 $U < 12$ stellar clusters and associations from
Bica et~al.\@ (1996) using the images presented
by Bell et~al.\@ (2002).
Symbol sizes in Fig.\ \ref{fig:bica} reflect a cluster's 
UV attenuation, as estimated from FUV/FIR.  The colors
of a single burst stellar population and dust screen reddening
vectors are also shown.

Clearly, {\it only} young, unattenuated clusters have
`blue' $\beta \sim -2$ values.  Redder, $-1 \la \beta \la 1$
clusters tend to be {\it either} relatively young but
attenuated (the clusters with $U - B \sim -0.8$, but redder $\beta$ 
values) {\it or}  older and dust free (the clusters 
with $U - B \sim 0$).  

The balance between dust and old stellar
population effects on the LMC's overall color
can be constrained by considering the 
fraction of the total $U < 12$ LMC cluster UV luminosity 
which each population represents.  The young, unattenuated clusters
represent 67\% of the summed $U < 12$ LMC cluster FUV 1500{\AA} 
luminosity.  The younger, attenuated
clusters represent 27\% of the FUV luminosity.  
The older, unattenuated clusters have only 6\% of the FUV 
luminosity.  

This result tentatively ascribes much of the 
observed `redness' of the LMC to dust effects:
older stellar populations tend to be UV-faint and do not 
affect the global $\beta$ estimate as significantly.  This 
interpretation is consistent with the detailed results of stellar
population modeling.  Leitherer et~al.\@ (1999) show that the {\it maximum} 
possible offset $\Delta\beta$ between young 
and older {\it star-forming} stellar 
populations is $\sim$0.5, as, to first order, stars that are bright
enough to affect the FUV luminosity of a galaxy with even a small
amount of ongoing SF have very blue $\beta$ values.
This lends weight to an interpretation of 
the redder $\beta$ values of normal galaxies mostly in 
terms of dust, with a small contribution from SF histories.

\section{Exploring alternatives to $\beta$} \label{sec:alt}

\begin{figure}[tbh]
\vspace{-0.1cm}
\hspace{1.0cm}
\epsfxsize=6.7cm
\epsfbox{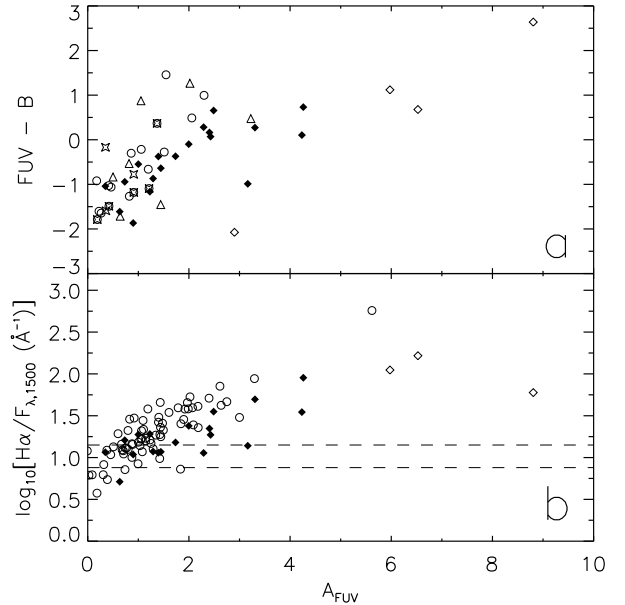}
\vspace{0.5cm}
\caption{\label{fig:alt} 
Alternatives to $\beta$.  Panel {\it a)} shows how FUV$ - B$ 
colors correlate with \afuvns. 
Symbols are as in Fig.\ \protect\ref{fig:beta}.
Panel {\it b)} shows \hans/\fuv against \afuvns.
Normal galaxies are shown as open circles,
starburst galaxies are shown as filled diamonds, and ULIRGs
are shown as open diamonds.
The two dashed lines depict the rough
dust-free value of \hans/\fuv (e.g., Kennicutt 1998,
Sullivan et~al.\@ 2000).
}
\end{figure}

Fig.\ \ref{fig:beta} demonstrates that it is difficult
to estimate UV attenuations
on the basis of UV colors alone.  For example, a galaxy with 
$\beta \sim -1$ could have zero attenuation (if it is a 
relatively quiescent galaxy), or many magnitudes of 
attenuation (if it is a ULIRG).  
In Fig.\ \ref{fig:alt}, I examine two possible 
alternatives to $\beta$ which use only rest-frame
UV and optical data (and are therefore more easily
accessible to researchers wishing to determine SF rates
at high redshift).  

It is conceivable that FUV$ - B$, because of its longer
wavelength range, may be more robust to dust
radiative transfer effects than the UV
spectral slope $\beta$ (but would suffer more
acutely from the effects of older stellar populations). 
In panel {\it a)} of Fig.\ \ref{fig:alt}, I show
1550{\AA} FUV$ - B$ colors against \afuv 
for my sample galaxies.  It is possible that 
FUV$ - B$ is a slightly more robust indicator than 
$\beta$, in terms of estimating \afuvns.  However, 
the scatter is enormous: at a given FUV$ - B$ color,
the range in \afuv is 3--5 magnitudes, or between 1 and 2
orders of magnitude.

Buat et~al.\@ (2002) suggest another potential extinction indicator:
\hans/\fuvns.  This indicator has the virtue that it is almost
independent of SF history (although see Sullivan et~al.\@ 2000); however, 
it does depend on the relative distribution of dust around 
\hii regions compared to the dust around OB associations 
(see, e.g., Bell et~al.\@ 2002).
In panel {\it b)} of Fig.\ \ref{fig:alt}, I show integrated
\hans/\fuv for a sample of normal galaxies from Bell \& Kennicutt (2001)
as open circles.
Matched aperture 
\hans/\fuv values for starburst galaxies 
(Calzetti et~al.\@ 1994), and for 
ULIRGs (Goldader et~al.\@ 2002; Wu et~al.\ 1998) are also shown.
In agreement with Buat et~al.\@ (2002), I find that 
there is a scattered correlation 
between \hans/\fuv and \afuvns; however, the scatter is a 
challenge to its usefulness.  For example, 
at \hans/\fuv$ \sim 10${\AA}$^{-1}$, $0\,{\rm mag} 
\la A_{\rm FUV} \la 3\,{\rm mag}$, and at \hans/\fuv$ \sim 100${\AA}$^{-1}$, 
$A_{\rm FUV}  \ga 3\,{\rm mag}$.  

Importantly, at a given \afuvns, 
starbursts and ULIRGs tend to have bluer $\beta$ values, bluer FUV$ - B$
colors, and lower \hans/\fuv than normal galaxies.
Thus, SF rates derived from rest frame UV data, even
analyzed in conjunction with rest frame optical data, suffer from 
{\it systematic} uncertainties of at least factors of a few.

\section{Conclusions} \label{sec:conc}

Seven independent UV experiments
demonstrate that quiescent, `normal' star-forming galaxies 
have substantially redder UV spectral slopes $\beta$ at
a given \afuv than starbursting galaxies.
Using spatially resolved data for the LMC, 
I argue that dust geometry and properties, coupled with 
a small contribution from older stellar populations, 
cause deviations from the
starburst galaxy $\beta$--\afuv correlation.
Neither rest frame UV-optical colors nor 
UV/\ha significantly help to constrain the UV attenuation.
Thus, SF rates estimated from rest-frame UV and optical data alone are
subject to large (factors of at least a few) systematic uncertainties
because of dust, which cannot be reliably corrected for
using only UV/optical diagnostics.

However, SF rates for high
$z$ galaxies derived from 
other wavelengths are also often subject to systematic
errors of this magnitude.  For example, sub-mm fluxes for high-redshift
star forming galaxies sample rest-frame $\sim 200${\micron}:
they must be converted to total IR flux assuming some dust spectrum.
Adelberger \& Steidel (2000) derive values of $\nu I_{\nu}$ at
200{\micron} vs.\ $L_{\rm FIR}$ of about 0.06 and 0.13 for 
ULIRGs and starbursts respectively.  Using Tuffs et~al.'s (2002) 
{\it ISO} photometry of star-forming galaxies 
at 170{\micron} as a constraint, I 
find that $\nu I_{\nu}$ at
$\sim$200{\micron} vs.\ $L_{\rm FIR}$ is roughly 0.2 for galaxies with warm
dust ($100/60{\micron} \sim 1$), growing to $\sim 0.6$
for galaxies with cold dust ($100/60{\micron} \sim 6$).
Thus, there is a factor of $\sim 10$
systematic error because of dust temperature which 
affects submm-derived SF rates (see also Dunne \& Eales 2001).
A similar limitation\adjustfinalcols affects radio-derived SF rates: because of
the mismatch in cosmic ray propagation and SF timescales, 
an order of magnitude scatter between radio flux and SF rate is
easily possible (Bressan et~al.\@ 2002).
Thus, when it comes to deriving SF rates for high-redshift galaxies
from data at almost any wavelength,
we are playing, at best, an order-of-magnitude game.

This work was supported by NASA grant NAG5-8426 and NSF grant AST-9900789.

\end{document}